\documentclass[11pt,onecolumn]{article}
\usepackage{a4wide}
\usepackage{mathrsfs}
\usepackage{latexsym,bm}
\usepackage{graphicx}
\usepackage{mdframed}
\usepackage{subfigure}
\usepackage{indentfirst}
\usepackage{slashed}
\usepackage{amsmath}
\usepackage{amssymb}
\usepackage{color}
\usepackage{hyperref}
\usepackage{epsfig}
\usepackage[titletoc]{appendix}
\usepackage{multirow}%
\usepackage{rotating}
\usepackage{cite}
\usepackage{ulem} 
\usepackage[top=1.5cm,bottom=1.9cm,left=2.5cm,right=2.5cm]{geometry}
\usepackage{tabularx}
\renewcommand{\arraystretch}{1.5}

\newcommand{\mL}{\mathcal{L}}

\newcommand{\bra}{\langle}
\newcommand{\ket}{\rangle}
\newcommand{\nn}{\nonumber}

\newcommand{\mT}{\mathcal{T}}

\newcommand{\uao}{U_A(1)}

\begin{document}
\thispagestyle{empty}
\title{ \Large \bf Light-flavor resonance dynamics in the $U(3)$ chiral theory}
\author{\small Zhi-Hui Guo     \\[0.3em] 
{ \small\it  Department of Physics and Hebei Advanced Thin Films Laboratory, } \\
{\small\it Hebei Normal University,  Shijiazhuang 050024, China}
}
\date{}

\maketitle

\begin{abstract} 
We review the recent developments on the light-flavor resonances in the $U(3)$ chiral effective filed theory.  
The spectral function sum rules and the semilocal duality in the scattering, which will be focus of this note, can provide us interesting and useful theoretical objects to bridge the hadron resonances in the intermediate energy region and the QCD behaviors in the asymptotic region. First the calculations of the meson-meson scattering amplitudes and factor factors are elaborated. The scalar spectral functions are then calculated in terms of the unitarized scalar form factors. The scalar and pseudoscalar spectral function sum rules in our study are found to be consistent with the asymptotic behavior of QCD in the chiral limit. The semilocal duality is found to be generally well satisfied, indicating the necessary cancellations of different contributions from different resonances indeed happen in the scattering amplitudes. The $N_C$ evolutions of the resonance poles, the ratios to quantify the semilocal duality and the spectral function integrals are also paid special attention to in this note. 
\end{abstract}

\section{Introduction}

Hadron resonances manifest themselves in the strongly interacting systems of the underlying hadronic states, such as the two- and many-body scattering processes, the form factors and the spectral functions of the hadrons, etc. The combination of the chiral effective field theory and the requirements of the unitarity and analyticity provides a reliable and powerful tool to systematically investigate the hadronic resonances appearing in various physical quantities. The methodology developed in the hadronic sector could also shed light on the exploring study of the possible resonances beyond the Standard Model that appear in the scattering processes of the $W$ and $Z$ bosons~\cite{bibchptew}.

The spontaneous $SU(3)_L\times SU(3)_R\to SU(3)_V$ chiral symmetry breaking of QCD leads to the eight pseudo-Nambu-Goldstone bosons (pNGBs), which can be identified as the eight light mesons $\pi, K$ and $\eta$. The small masses of the pNGBs are caused by the explicit breaking of the chiral symmetry from the light-flavor quark masses. Chiral perturbation theory ($\chi$PT)~\cite{Weinberg:1978kz,Gasser:1983yg,Gasser:1984gg}, as the first well-established effective theory of QCD, has been extensively demonstrated to be powerful to describe the physical processes involving the pNGBs $\pi, K$ and $\eta$. 

Another important property of QCD at low energy is the $U_A(1)$ anomaly arising from the strong interactions, which is believed to be responsible for the large mass of the singlet $\eta_0$, that gives the most important component to the physical $\eta'$ state. Due to the $\uao$ anomaly effect, $M_0$, the mass of the singlet $\eta_0$, does not vanish and keeps a large value around 1~GeV even in the chiral limit. The appearance of the new scale $M_0$ breaks down the conventional chiral power-counting scheme, which relies on the perturbative expansions of the external momenta and light meson masses. One way to systematically include the $\eta'$ state in $\chi$PT is the large $N_C$ framework, being $N_C$ the number of colors in QCD. According to the large $N_C$ QCD~\cite{largenc}, the quark loops, which are responsible for the QCD $U_A(1)$ anomaly~\cite{ua1anomaly}, are $1/N_C$ suppressed, implying that in the $N_C\to \infty$ limit the QCD $\uao$ anomaly would disappear and the singlet $\eta_0$ would become a pNGB in the chiral limit~\cite{ua1nc}, as the $\pi, K$ and $\eta$. In this framework, the leading order mass squared of the $\eta_0$, $M_0^2$, scales as $1/N_C$ when $N_C\to \infty$. Based on this argument, the triple $\delta$ expansion scheme, i.e. $O(\delta) \sim O(1/N_C) \sim O(p^2) \sim O(m_q)$, is proposed to simultaneously study the $\pi, K, \eta$ and $\eta'$ in $\chi$PT, which is also referred as $U(3)$ $\chi$PT in literature~\cite{HerreraSiklody:1996pm,Kaiser:2000gs}. By taking both the chiral and large $N_C$ limits, the dynamical degrees of freedom of the very low energy QCD would be the nonet $\pi, K, \eta_8$ and $\eta_0$. From this point of view, specially when one attempts to probe the $N_C$ behaviors of various hadron resonances, the $U(3)$ $\chi$PT offers a better motivated theoretical framework to study the resonance properties than the conventional $SU(3)$ case~\cite{Gasser:1984gg}.

During the last decade, important progresses on the $U(3)$ $\chi$PT within the $\delta$ expansion scheme have been made, including the one-loop calculation of all the two-meson scattering amplitudes, scalar and pseudoscalar form factors and the study of the $N_C$ behaviors of the various light-flavor scalar and vector resonances~\cite{Guo:2011pa,Guo:2012ym,Guo:2012yt,Guo:2016zep}. Special attention has been paid to the light-flavor resonance dynamics in the scalar and pseudoscalar spectral sum rules, and the semilocal duality from the meson-meson scattering. In this note, we first briefly introduce the theoretical formalism and then discuss the key findings of the recent $U(3)$ $\chi$PT  developments.

\section{The theoretical setups of the $U(3)$ chiral theory}\label{sec.theorysetup}

At leading order (LO) in the $\delta$ expansion, the $U(3)$ Lagrangian consists of three independent terms
\begin{eqnarray}\label{eq.laglo}
\mL^{(0)}= \frac{ F^2}{4}\bra u_\mu u^\mu \ket+
\frac{F^2}{4}\bra \chi_+ \ket
+ \frac{F^2}{12}M_0^2 X^2 \,,
\end{eqnarray}
where the basic chiral operators take the form
\begin{eqnarray}\label{defbb}
&& U =  u^2 = e^{i\frac{ \sqrt2\Phi}{ F}}\,, \qquad \chi = 2 B (s + i p) \,, \nn \\&& 
X= \ln{(\det U)}\,, \quad \chi_\pm  = u^\dagger  \chi u^\dagger  \pm  u \chi^\dagger  u \,,  \quad u_\mu = i u^\dagger  D_\mu U u^\dagger \,,   \nn \\ &&
D_\mu U \, =\, \partial_\mu U - i (v_\mu + a_\mu) U\, + i U  (v_\mu - a_\mu) \,, 
\end{eqnarray}
and the contents of the pNGBs in the $U(3)$ case are given by
\begin{equation}\label{eq.phi9}
\Phi \,=\, \left( \begin{array}{ccc}
\frac{1}{\sqrt{2}} \pi^0+\frac{1}{\sqrt{6}}\eta_8+\frac{1}{\sqrt{3}} \eta_0 & \pi^+ & K^+ \\ \pi^- &
\frac{-1}{\sqrt{2}} \pi^0+\frac{1}{\sqrt{6}}\eta_8+\frac{1}{\sqrt{3}} \eta_0   & K^0 \\  K^- & \overline{K}^0 &
\frac{-2}{\sqrt{6}}\eta_8+\frac{1}{\sqrt{3}} \eta_0
\end{array} \right)\,.
\end{equation}
Here $F$ stands for the pion decay constant at LO, with the physical normalization $F_\pi=92.1$~MeV. $v_\mu, a_\mu, p$ and $s$ represent the vector, axial-vector, pseudoscalar and scalar external sources, respectively. By taking the vacuum expectation values of the scalar source as $s={\rm diag}(m_u,m_d,m_s)$, being $m_{q=u,d,s}$ the light quark masses, one can implement the explicit chiral symmetry breaking in the same way as that in QCD. It is noted that we work in the isospin symmetric situation throughout, that is to take $m_u=m_d=\hat{m}$. The quantity $B$ is proportional to the quark condensate via $\bra 0| \bar{q}^{a}q^{b} |0\ket = - B\,F^2 \delta^{ab}$. The last term in Eq.~\eqref{eq.laglo} introduces the QCD $\uao$ anomaly effect and gives the $\eta_0$ the LO mass $M_0$.

Generally, the higher-order operators in the effective field theory incorporates the higher energy dynamics. So that, apart from introducing the higher order local operators, another approach to take into account the higher order effects is to explicitly include dynamical degrees of freedom beyond low energy ones. One of such successful attempts along the line of this research is the resonance chiral theory (R$\chi$T), which explicitly includes the light-flavor vector, scalar and pseudoscalar resonances in its construction of the Lagrangians~\cite{Ecker:1988te}. Later on R$\chi$T also turns out to be very useful for the phenomenological study of the physical processes involving  resonances~\cite{rxtpheno}. Although the rigorous  generalization of the R$\chi$T into the loop calculation still faces problems, the large $N_C$ QCD argument provides useful guidelines for the construction of the R$\chi$T operators~\cite{Cirigliano:2006hb}. The relevant ones that enter the meson-meson scattering, the scalar and pseudoscalar form factors in our study are analyzed in detail in Refs.~\cite{Guo:2011pa,Guo:2012ym,Guo:2012yt}. The R$\chi$T Lagrangian that describes the interactions between the pNGBs and the vector resonances reads~\cite{Ecker:1988te} 
\begin{eqnarray}\label{eq.lagvector}
\mL_{V}= \frac{i G_V}{2\sqrt2}\langle V_{\mu\nu}[ u^\mu, u^\nu]\rangle \,,
\end{eqnarray}
and the one describing the interactions between the pNGBs and the scalar resonances is given by 
\begin{align}\label{eq.lagscalar}
\mL_{S}&= c_d\bra S_8 u_\mu u^\mu \ket + c_m \bra S_8 \chi_+ \ket
 + \widetilde{c}_d S_1 \bra u_\mu u^\mu \ket
+ \widetilde{c}_m  S_1 \bra  \chi_+ \ket \,.
\end{align} 
The relevant Lagrangian involving the pseudoscalar resonances is 
\begin{eqnarray}\label{eq.lagpscalar}
\mL_{P} = i d_m \bra P_8 \chi_- \ket + i \widetilde{d}_m  P_1 \bra \chi_- \ket\,.
\end{eqnarray}
For the definitions of the explicit matrix contents of the vector nonet $V$, scalar octet $S_8$ and scalar singlet $S_1$, we refer to Ref.~\cite{Guo:2011pa} and references therein for details. The matrix contents of the pseudoscalar resonances share the same flavor structures of the pNGBs in Eq.~\eqref{eq.phi9}. Since we focus on the resonance dynamics in the study of the meson-meson scattering and the form factors, the R$\chi$T Lagrangians in Eqs.~\eqref{eq.lagvector}, \eqref{eq.lagscalar} and \eqref{eq.lagpscalar} are used in the calculation. By integrating out the resonances, one can get the higher-order local operators, most of which are the ones surviving in the large $N_C$ limit. In the $\delta$ expansion, there are two next-to-leading order (NLO) pure $U(3)$ local operators, namely 
\begin{eqnarray}\label{eq.laglam}
\mL^{(1)} &=&-\frac{F^2\, \Lambda_1}{12}   \partial^\mu X \partial_\mu X  -\frac{F^2\, \Lambda_2}{12} X \bra \chi_- \ket\,,
\end{eqnarray}
in the sense that they do not appear in the $SU(3)$ $\chi$PT. These two operators can not be generated from the resonance Lagrangians in Eqs.~\eqref{eq.lagvector},~\eqref{eq.lagscalar} and~\eqref{eq.lagpscalar}. So we explicitly include the $\Lambda_1$ and $\Lambda_2$ operators in the phenomenological study. Moreover, we also include a remnant part of the $L_8$ term to account for the large uncertainties of the pseudoscalar resonances~\cite{Guo:2012ym,Guo:2012yt}.

\begin{figure}[htbp]
\centering
\includegraphics[width=0.95\textwidth,angle=-0]{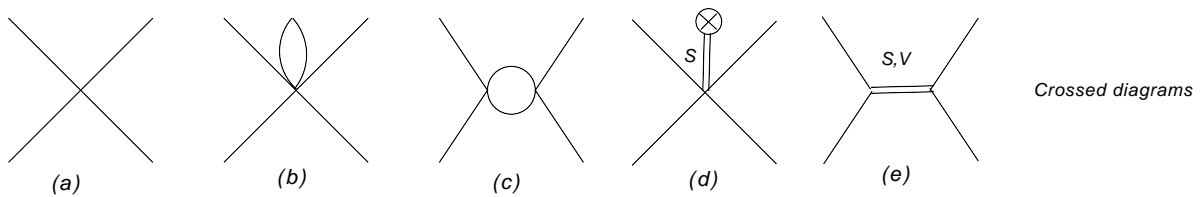} 
  \caption{ 1PI Feynman diagrams for the meson-meson scattering up to one loop with explicit tree-level exchanges.   }
   \label{fig.fyndiagsc}
\end{figure}

\begin{figure}[htbp]
   \centering
      \begin{minipage}[t]{0.45\textwidth}
         \centering
   \includegraphics[width=0.95\textwidth,angle=-0]{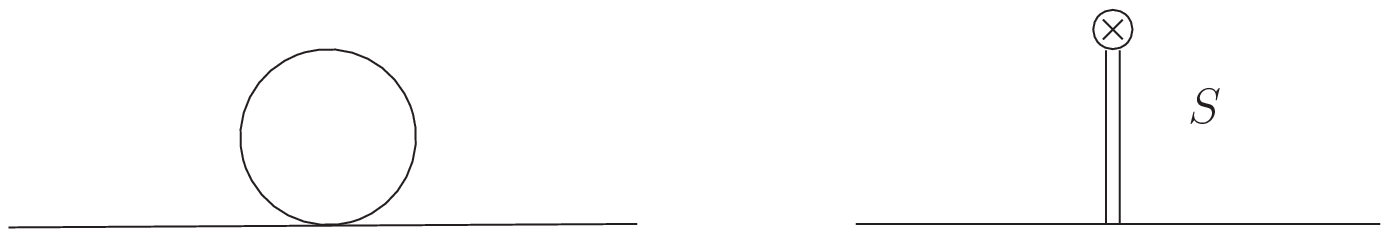} 
   \end{minipage}
     \begin{minipage}[t]{0.45\textwidth}
    \centering
   \includegraphics[width=0.95\textwidth,angle=-0]{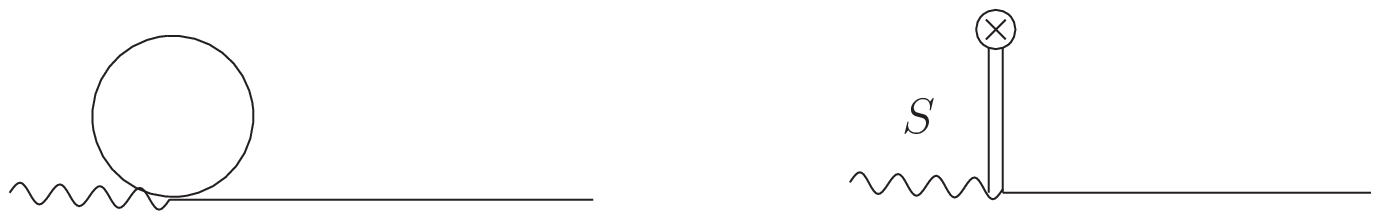} 
   \end{minipage}
  \caption{ Feynman diagrams for the one-loop self-energy (the left two diagrams) and the decay constant (the right two diagrams) for the light pseudoscalar meson.   }
   \label{fig.fyndiagsefp}
\end{figure}

\begin{figure}[htbp]
   \centering
      \begin{minipage}[t]{0.45\textwidth}
         \centering
   \begin{mdframed}
\includegraphics[width=0.99\textwidth,angle=-0]{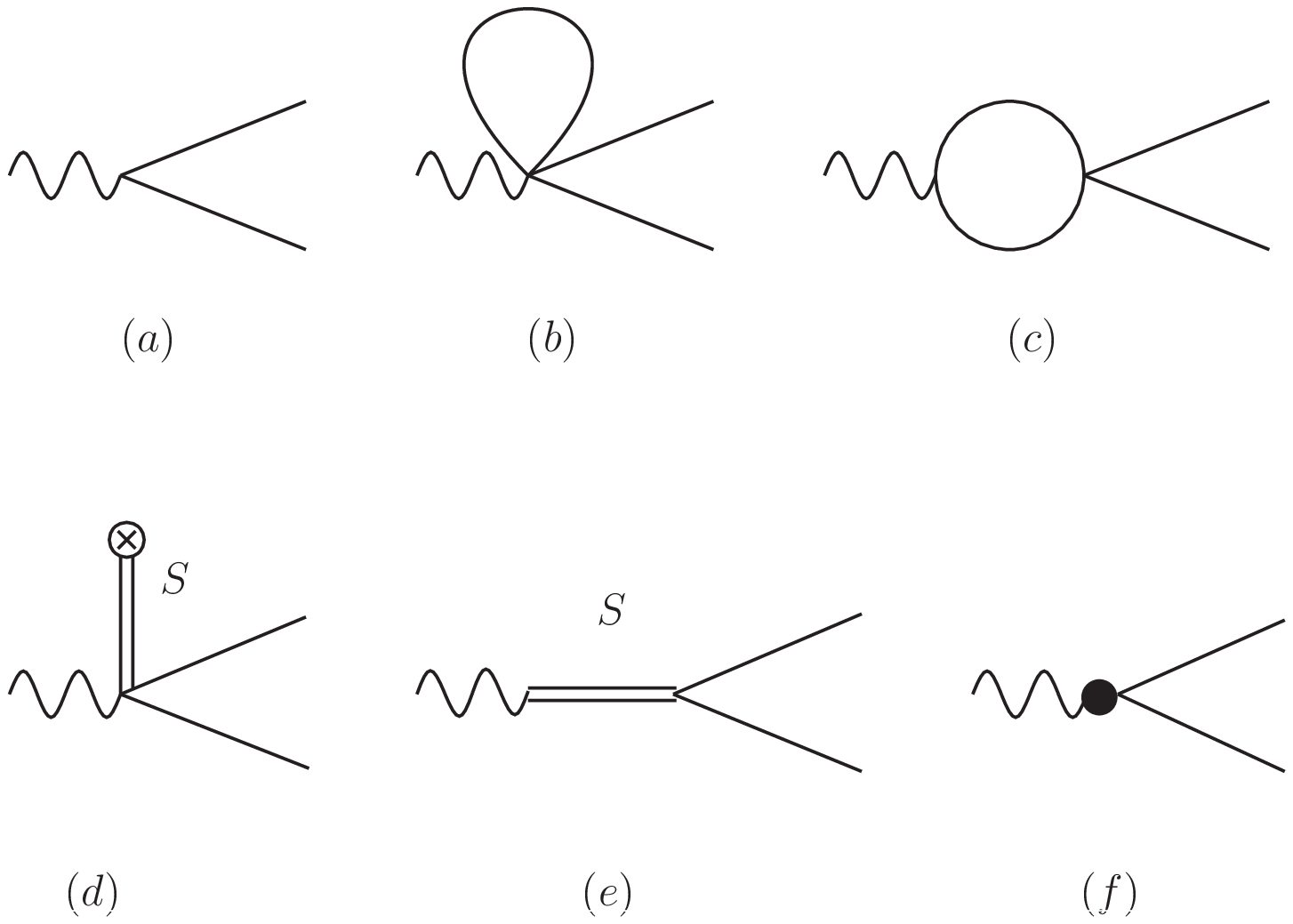} 
    \end{mdframed}
   \end{minipage}
     \begin{minipage}[t]{0.45\textwidth}
    \centering
      \begin{mdframed}
\includegraphics[width=0.99\textwidth,angle=-0]{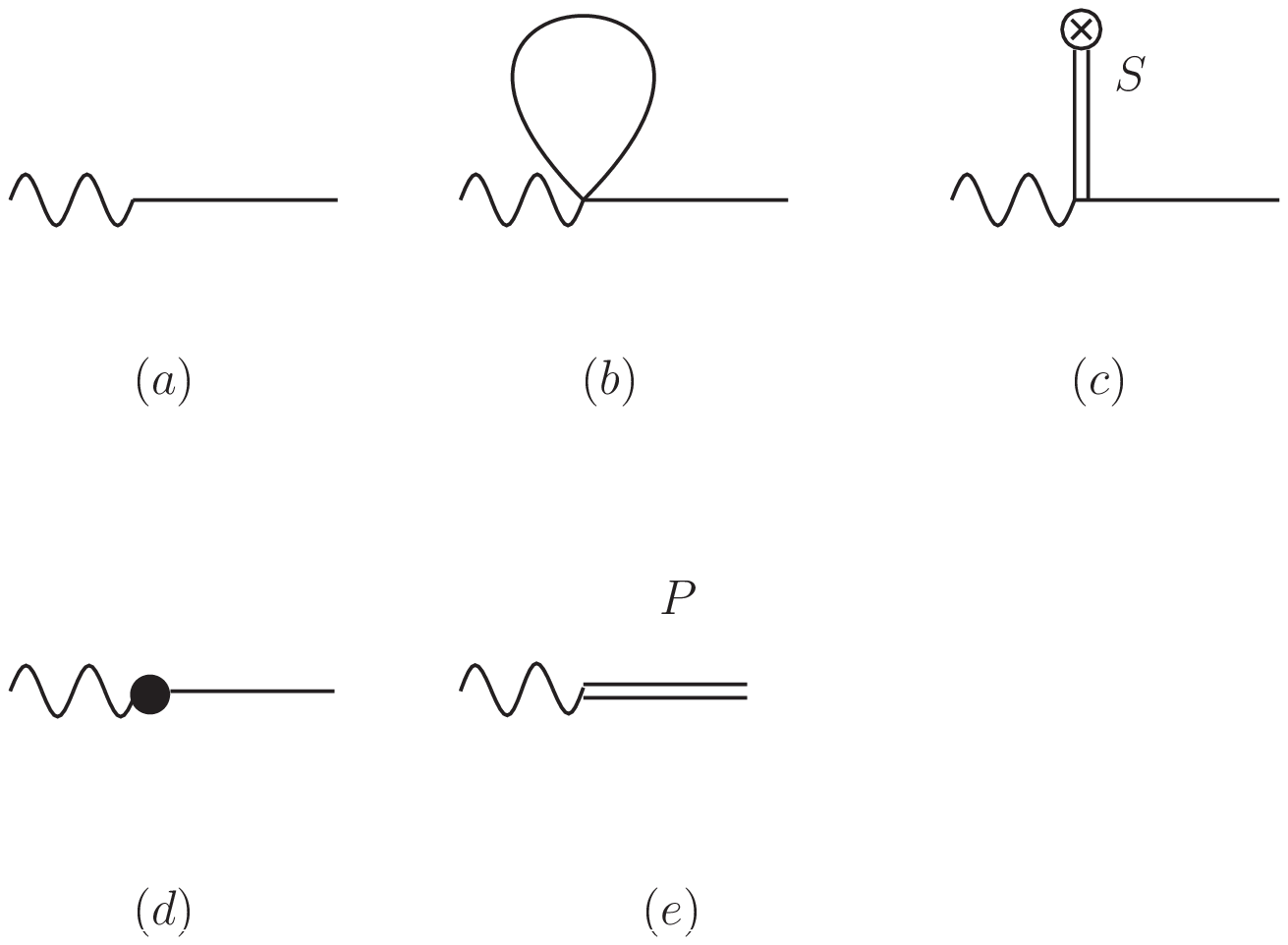} 
      \end{mdframed}
   \end{minipage}
  \caption{ Feynman diagrams for the scalar (left panel) and pseudoscalar (right panel) form factors.   }
   \label{fig.fyndiagff}
\end{figure}

The complete one-loop two-pNGB scattering amplitudes involving the $\pi, K, \eta$ and $\eta'$ with tree-level resonance exchanges in the $U(3)$ chiral theory have been calculated in Ref.~\cite{Guo:2011pa}. The one-particle-irreducible (1PI) Feynman diagrams of the two-pNGB scattering amplitudes are illustrated in Fig.~\ref{fig.fyndiagsc}. The one-loop diagrams of the self-energies and the decay constants of the light mesons are given in Fig.~\ref{fig.fyndiagsefp}. The 1PI one-loop diagrams with tree-level resonance exchanges of the two-pNGB scalar form factors and the one-pNGB pseudoscalar form factors are shown in Fig.~\ref{fig.fyndiagff} and they are calculated in Refs.~\cite{Guo:2012ym,Guo:2012yt}. Within the $\delta$ expansion scheme, the conventional dimensional regularization method that is used in the $SU(2)$ and $SU(3)$ $\chi$PT~\cite{Gasser:1983yg,Gasser:1984gg}, is still valid in the $U(3)$ case, since the new scale $M_0$, behaves as $1/N_C$ in the large $N_C$ limit. While by treating the $\eta_0$ as a heavy field, the dimensional regularization will break the well-established chiral power-counting rule and different regularization methods have been correspondingly suggested to study the $\eta'$ in Refs.~\cite{Beisert:2001qb,Beisert:2002ad}.

To incorporate the meson-meson nonperturbative strong interactions at the resonance energy region, the perturbative calculations elaborated in Figs.~\ref{fig.fyndiagsc} and \ref{fig.fyndiagff}, even after the explicit inclusion of the bare resonance exchanges, are not enough, because the chiral interactions between the meson pairs could increase rapidly when the energies lie above the two-meson thresholds. One efficient way to include such nonperturbative effects is to perform the unitarization of the amplitudes, and  there are vast literatures on this subject, see recent comprehensive and pedagogical reviews in Refs.~\cite{Oller:2019opk,Oller:2020guq} and references therein for further details. The basic unitarization formalism for the two-body partial-wave scattering amplitudes that we use is an approximated version of the $N/D$ method~\cite{Oller:1998zr} 
\begin{eqnarray} \label{eq.ut}
\mT_{IJ}(s) = \dfrac{N_{IJ}(s)}{1 - N_{IJ}(s)\,G_{IJ}(s)}\,, 
\end{eqnarray}
where the subscripts $IJ$ denote the quantum numbers of the isospin and angular momentum. We will omit the subscripts $IJ$ in the next discussions for simplicity. By construction, the quantity $N(s)$ only contains the crossed-channel contributions (including the local contact terms as well) and the function $G(s)$ includes the right-hand cut contributions. Above the threshold the unitarity of the S matrix determines the imaginary part of the $G(s)$ function 
\begin{equation}\label{eq.img}
{\rm Im}G(s) = \rho(s) \, \equiv  \dfrac{q(s)}{8\pi\sqrt{s}} \,,  \,\,\, (s>s_{\rm th})\,
\end{equation}
where $s_{\rm th}=(m_1+m_2)^2$ stands for the threshold of the two particles with masses $m_1$ and $m_2$, and the center of mass (CM) three momentum takes the form
\begin{equation}\label{eq.q3}
q(s) = \dfrac{\sqrt{[s-(m_1+m_2)^2][s-(m_1-m_2)^2]}}{2\sqrt{s}}\,.
\end{equation} 
The K-matrix unitarization method includes only the imaginary part of the $G(s)$ function in its construction of the unitarized scattering amplitudes. Clearly by taking only the imaginary part of the $G(s)$ the analyticity is not preserved. One can improve the K-matrix description by using a once subtracted dispersion relation to include the real part of the $G(s)$ function. It turns out that  by using the dimensional regularization method to calculate the one-loop two-point function we can also obtain the same result for the $G(s)$ function by replacing the divergent term with a free subtraction constant. The explicit expression takes the form~\cite{Oller:1998zr} 
\begin{eqnarray}\label{eq.gfunc}
G(s)^{\rm DR} &=& -\frac{1}{16\pi^2}\left[ a(\mu^2) + \log\frac{m_2^2}{\mu^2}-x_+\log\frac{x_+-1}{x_+}
-x_-\log\frac{x_--1}{x_-} \right]\,, 
\end{eqnarray}
where $\mu$ is the regularization scale and will be set to $\mu=770$~MeV throughout, and $x_\pm$ are given by 
\begin{equation}
 x_\pm =\frac{s+m_1^2-m_2^2}{2s}\pm \frac{q(s)}{\sqrt{s}}\,.
\end{equation}
Notice that due to the inclusion of the minus sign of the $G(s)$ function, comparing with the definition in Refs.~\cite{Guo:2011pa,Guo:2012ym,Guo:2012yt}, the positive sign in the denominator becomes the minus sign in Eq.~\eqref{eq.ut}. The corresponding changes should also apply in the following discussions. By matching the unitarized amplitudes of Eq.~\eqref{eq.ut} and the perturbative chiral amplitudes order by order~\cite{Oller:2000fj}, one can obtain the expression for the $N(s)$ function 
\begin{eqnarray}\label{eq.defn}
N_{IJ}(s) = T_{IJ}(s)^{\rm LO+Res+Loop} - T_{IJ}(s)^{\rm LO}\,G_{IJ}(s)\,T_{IJ}(s)^{\rm LO}\,,
\end{eqnarray}
where $T_{IJ}(s)$ stand for the partial-wave projections of the perturbative chiral amplitudes. The explicit expressions for the leading order (LO), resonance exchanges (Res) and loop diagrams (Loop) are given in Ref.~\cite{Guo:2011pa}. 

For the scalar form factors of the two-meson states, we use a similar unitarization method to resum the nonperturbative strong interactions between the two mesons~\cite{Guo:2012ym,Guo:2012yt}
\begin{eqnarray}
\mathcal{F}_{I}(s)= \frac{R_{I}}{1-N_{IJ}(s)\,G_{IJ}(s)}\,,
\end{eqnarray}
where $N_{IJ}(s)$ is given by Eq.~\eqref{eq.defn} and $R_{I}(s)$ can be obtained by matching the unitarized form factor $\mathcal{F}(s)$ and the perturbative chiral results
\begin{eqnarray}
R_{I}(s) = F_{I}(s)^{\rm LO+Res+Loop} - N_{IJ}(s)^{\rm LO}\,G_{IJ}(s)\,F_{I}(s)^{\rm LO}\,.
\end{eqnarray}
$F_{I}(s)$ stands for the scalar form factors from the perturbative chiral calculation with definite isospin number $I$. The explicit expressions for the chiral perturbative scalar form factors are given in Ref.~\cite{Guo:2012yt}.
In the coupled-channel case with $n$ channels, one should understand the functions $N_{IJ}(s)$ and $G_{IJ}(s)$ as $n\times n$ matrices, and $R_{I}(s)$ as an $n$-row vector.

\section{Resonance dynamics in the meson-meson scattering, form factor and spectral functions}

In order to fix the unknown parameters, we have fitted a large amount of experimental and lattice data. The experimental data include the phase shifts and inelasticities from the $S$- and $P$-wave $\pi\pi \to \pi\pi$,  $\pi\pi \to K\bar{K}$ and $\pi K \to \pi K$ scattering processes. Regarding the $\pi\eta$ case, the experimental measurement on this scattering process is still not available and there are only various $\pi\eta$ event distributions, which would require more theoretical inputs apart from the $\pi\eta$ scattering parameters. It turns out that the uncertainties of the $\pi\eta$ scattering amplitudes obtained from the fits to the event distributions alone are quite large~\cite{Guo:2011pa,Guo:2012ym,Guo:2012yt}. This motivates us to include the lattice finite-volume spectra from the Hadron Spectrum Collaboration~\cite{Dudek:2016cru} to further constrain the $\pi\eta$, $K\bar{K}$ and $\pi\eta'$ coupled-channel scattering amplitudes~\cite{Guo:2016zep}. Interested readers are recommended to go through Refs.~\cite{Guo:2011pa,Guo:2012yt,Guo:2016zep} for details of the fit results. Next we elaborate the resonance dynamics in the various physical processes. 

\begin{table}[htbp]
\renewcommand{\tabcolsep}{0.05cm}
\renewcommand{\arraystretch}{1.2}
\begin{center}
{\small 
\begin{tabular}{|c|c|c| |c|c|c|}
\hline
R & M (MeV) & $\Gamma/2$ (MeV) & R & M (MeV) & $\Gamma/2$ (MeV) 
\\
[0.1cm] 
\hline
\hline
$f_0(500)$& $442^{+4}_{-4}$ & $246^{+7}_{-5}$  &  $\rho(770)$& $760^{+7}_{-5}$ & $71^{+4}_{-5}$  
\\[0.1cm]   
\hline 
$f_0(980)$& $978^{+17}_{-11}$ & $29^{+9}_{-11}$ &  $K^*(892)$& $892^{+5}_{-7}$ & $25^{+2}_{-2}$ 
\\[0.1cm]  \hline 
$f_0(1370)$& $1360^{+80}_{-60}$ & $170^{+55}_{-55}$ & $\phi(1020)$& $1019.1^{+0.5}_{-0.6}$ & $1.9^{+0.1}_{-0.1}$ 
\\[0.1cm]  
\hline
$K^*_0(800)$& $643^{+75}_{-30}$ & $303^{+25}_{-75}$ &  $a_0(980)$ & $1019^{+22}_{-8}$ & $24^{+57}_{-17}$ 
\\[0.1cm] 
 \hline  
$K^*_0(1430)$&  $1482^{+55}_{-110}$ & $132^{+40}_{-90}$ & $a_0(1450)$ & $1397^{+40}_{-27}$ & $62^{+79}_{-8}$ 
 \\[0.1cm]  
\hline
\end{tabular}
}
 \caption{ The masses and the half widths of the resonances appearing in the meson-meson scattering. The resonance poles of the $a_0(980)$ and $a_0(1450)$ are determined by fitting simultaneously the experimental $\pi\eta$ event distributions, the cross sections of $\gamma\gamma \to \pi\eta$ and also the lattice finite-volume spectra with the NLO chiral amplitudes~\cite{Guo:2016zep}. The other resonance poles are determined by fitting the experimental phase shifts and inelasticities~\cite{Guo:2012yt}.    }  
\label{tab:pole}
\end{center}
\end{table}

\begin{figure}[htbp]
\centering
\includegraphics[width=0.95\textwidth,angle=-0]{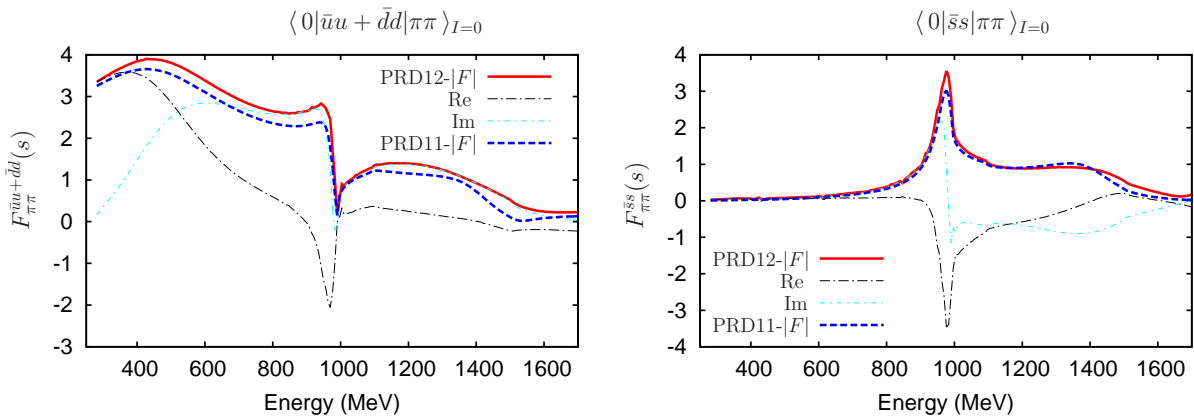} 
  \caption{ Results for the scalar $\pi\pi$ form factors. The curves labeled as PRD11 and PRD12 are obtained by using the parameters from Ref.~\cite{Guo:2011pa} and Ref.~\cite{Guo:2012yt}, respectively.   }
   \label{fig.sffpipi}
\end{figure}

The resonance contents are briefly summarized in Table~\ref{tab:pole}, where both the mass (real part) and the half width (imaginary part) for each pole are given. It is interesting to dissect the roles of these resonance poles that are played in the various physical quantities, such as the form factors, spectral functions and semilocal duality from the scattering. In Fig.~\ref{fig.sffpipi}, we show two different types of $\pi\pi$ scalar form factors with the scalar densities of $\bar{u}u+\bar{d}d$ and $\bar{s}s$. The real parts, imaginary parts and the magnitudes from the two similar fit results in Refs.~\cite{Guo:2011pa,Guo:2012yt} are shown together in Fig.~\ref{fig.sffpipi}. It is clear that the $f_0(500)$ or the $\sigma$ resonance should be responsible for the low energy bump around $0.5$~GeV in the scalar form factor $F_{\pi\pi}^{\bar{u}u+\bar{d}d}(s)$, which is defined as 
\begin{eqnarray}\label{eq.defsffuudd}
B\,F_{\pi\pi}^{\bar{u}u+\bar{d}d}(s) = \bra 0|\, \bar{u} u + \bar{d} d \,|\,(\pi\pi)_{I=0} \,\ket\,. 
\end{eqnarray}
In contrast the broad $\sigma$ resonance barely contributes to the scalar form factor $F_{\pi\pi}^{\bar{s}s}(s)$, which is defined as
\begin{eqnarray}\label{eq.defsffss}
B\,F_{\pi\pi}^{\bar{s}s}(s) = \bra 0|\, \bar{s}s \,|\,(\pi\pi)_{I=0} \,\ket\,.
\end{eqnarray}
Interestingly, the $f_0(980)$ manifests itself as a dip in the $F_{\pi\pi}^{\bar{u}u+\bar{d}d}(s)$, but shows up as a peak in the $F_{\pi\pi}^{\bar{s}s}(s)$. While in the energy region above 1~GeV, we do not see any narrow structure appearing in both types of scalar form factors of $\pi\pi$. In this way, one can discern the role of the higher mass resonance $f_0(1370)$ played in the $\pi\pi$ scalar form factors. Other types of strangeness conserving two-meson scalar form factors defined as $\bra 0|\, \bar{q} \lambda_a q \,|\,PQ \,\ket$ are also calculated, with $\lambda_a$ the Gell-Mann matrices.

\begin{figure}[htbp]
\centering
\includegraphics[width=0.95\textwidth,angle=-0]{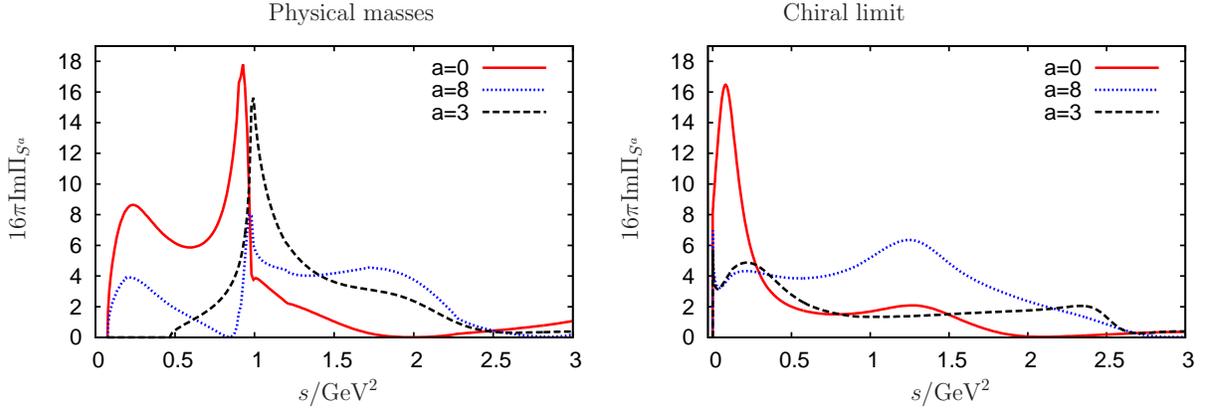} 
  \caption{ The scalar spectral functions with the definitions in Eq.~\eqref{eq.impi}. The left and right panels give the results with the parameters determined in Ref.~\cite{Guo:2012yt} by taking the physical masses and the chiral limit, respectively. }
   \label{fig.sffimpi}
\end{figure}

The scalar spectral function, i.e. the imaginary part of the correlating two-point scalar-density function, can be given by the scalar form factors via 
\begin{eqnarray}\label{eq.impi}
{\rm Im} \Pi_{S^{a}}(s) = \sum_i \rho_i(s) |F^{a}_i(s)|^2 \theta(s-s_i^{\rm th})\,,
\end{eqnarray}
where $i$ runs over the relevant two-meson channels, $s_i^{\rm th}$ stands for the threshold of the $i$th channel, $\theta(x)$ is the standard Heaviside step function and $\rho_i(s)$ is the kinematical factors defined in Eq.~\eqref{eq.img}. In Fig.~\ref{fig.sffimpi}, the scalar spectral functions of $a=0,3,8$ calculated with the parameters given in Refs.~\cite{Guo:2011pa,Guo:2012yt} are shown. 

In the chiral limit the scalar and pseudoscalar spectral functions follow a set of sum rules~\cite{speclsumrule}
\begin{eqnarray}
\int_0^{\infty} \big[ {\rm Im} \Pi_{X}(s) - {\rm Im} \Pi_{X'}(s) \big] ds = 0 \,,
\end{eqnarray}
with $X$ or $X'$ corresponding to either scalar or pseudoscalar light-flavor quark densities $\bar{q}\lambda_a q$ or $i\bar{q}\lambda_a \gamma_5 q$. These scalar and pseudoscalar spectral sum rules offer an important tool to study the complicated scalar resonance dynamics. It is advisory to split the integrals of the above sum rules into the nonperturbative and perturbative parts as
\begin{eqnarray}\label{eq.specint}
 \int_0^{s_0}\big[ {\rm Im} \Pi_{X}(s) - {\rm Im} \Pi_{X'}(s) \big]ds + \int_{s_0}^{\infty}\big[ {\rm Im} \Pi_{X}(s) - {\rm Im} \Pi_{X'}(s) \big] ds = 0 \,,
\end{eqnarray}
where the nonperturbative integrands in the $0$ and $s_0$ region can be evaluated via Eq.~\eqref{eq.impi} and the operator product expansion (OPE) can be used to calculate the perturbative parts above $s_0$. According to the QCD OPE calculation~\cite{speclsumrule}, in the high energy perturbative region different types of the spectral integrals with $X$ or $X'$ $=S^{a}, P^{b}$ are equal in the chiral limit, which implies that the second integrals in Eq.~\eqref{eq.specint} are always zero in the chiral limit. This reduces the discussions of the spectral sum rules to the nonperturbative integrals in Eq.~\eqref{eq.specint}. Since there is lack of a rigorous criteria to set the separation scale $s_0$, we have chosen three different values for $s_0$, namely 2.5, 3.0 and 3.5~GeV$^2$, in our study. In practice, we find that the uncertainty caused by the ambiguity of $s_0$ is not that big, because the form factors and the spectral functions in our method tend to vanish in the high energy region, or at least they approach to rather small values in magnitudes, which can be clearly seen in Fig.~\ref{fig.sffimpi}. According to the curves in the figure, it seems that the broad $\sigma$ resonance contributes to both the spectral functions with $a=0$ and $a=8$, i.e. the isocalar $SU(3)$ singlet and octet currents, respectively. To be more specific, the height of the bump around the $\sigma$ resonance region in the isocalar singlet spectral function is around twice as that in the octet case. Similarly the $f_0(980)$ peak in the $a=0$ case is also higher than that in the $a=8$ case. While for the $f_0(1370)$ resonance, it dominantly contributes to the isocalar octet spectral function. For the isovector spectral function with $a=3$, both $a_0(980)$ and $a_0(1450)$ will contribute. Based on the fit results from Refs.~\cite{Guo:2011pa,Guo:2012yt}, which are obtained by only including the $\pi\eta$ event distributions to constrain the $\pi\eta$ scattering, we conclude that the $a_0(980)$ peak is much more prominent than the one from the $a_0(1450)$. Nevertheless there will be large uncertainties from those determinations. It will be interesting to make an analysis of the $\pi\eta$ form factor and to further explore its phenomenological application in a future work. 

For the pseudoscalar spectral function, we will only consider the single-meson contributions, so that it is simply given by the $\delta$ functions
\begin{eqnarray}
 {\rm Im} \Pi_{P^a}(s)= \sum_k \pi \delta(s-m_{P_k}^2)|H_k^a(s)|^2\,,
\end{eqnarray}
where $k$ runs over the proper intermediate pNGBs and the one-meson pseudoscalar form factor $H_k^a$ is defined as 
\begin{eqnarray}
B\,H_k^{a}(s)=\bra 0 | i \bar{q} \lambda_a \gamma_5 q |P_k\ket\,. 
\end{eqnarray} 
The one-meson pseudoscalar form factors are calculated up to one-loop level, see the relevant Feynman diagrams in the right panel of Fig.~\ref{fig.fyndiagff}. 

In order to quantity the fulfillment of the spectral sum rules in Eq.~\eqref{eq.specint},  several quantities based on the nonperturbative spectral integrals are proposed
\begin{eqnarray}
\overline{W} &=& \frac{\sum_n W_n}{3\times 6}\,, \\
\sigma_W &=& \sqrt{\sum_n\frac{(W_n-\overline{W})^2}{3\times6 -1}}\,, 
\end{eqnarray}
with 
\begin{eqnarray}\label{eq.defw}
W_n = 16\pi  \int_0^{s_0} {\rm Im} \Pi_{n}(s)\, ds\,,\,\, \big( n=\{S^{0,3,8},P^{0,3,8}\} \big)\,.
\end{eqnarray}
The value $\sigma_W/\overline{W}$ can be interpreted as a parameter to judge at which level the spectral sum rules are satisfied. 
The results by taking the fit parameters from Ref.~\cite{Guo:2012yt} and the physical masses for the mesons are  
\begin{eqnarray}
 \overline{W}=9.0\,,\quad \sigma_W= 1.5\,\, \Rightarrow \,\, \frac{\sigma_W}{\overline{W}} = 0.16\,,
\end{eqnarray}
which implies that the scalar and pseudoscalar spectral sum rules in Eq.~\eqref{eq.specint} are only violated at the level  around $15\%$. However, the violation of such sum rules is more severe when taking the fit parameters from Ref.~\cite{Guo:2011pa}, which can reach around $30\%$. We have also tried to perform the chiral extrapolation of the spectral functions to the chiral limit case, which introduces more uncertainties due to the less controlled extrapolating behaviors of the subtraction constants in the unitarized amplitudes~\cite{Guo:2012ym,Guo:2012yt}.

The scalar and pseudoscalar spectral sum rules discussed previously enable us to discern the underlying relations of the scalar and pseudoscalar mesons. On the other hand, the semilocal or the average duality from the meson-meson scattering provides another interesting theoretical framework to study the possible relations between the scalar and the vector resonances. The semilocal/average duality here refers to the relations between the hadronic system and the Regge theory in the fixed-$t$ scattering amplitudes. The key object is given by
\begin{eqnarray}\label{eq.defsd}
 \int_{\nu_1}^{\nu_2} \nu^{-n}{\rm Im} T_{\rm t,Regge}^{I}(\nu,t) d\nu =  \int_{\nu_1}^{\nu_2} \nu^{-n}{\rm Im} T_{\rm t,Hadron}^{I}(\nu,t) d\nu\,,
\end{eqnarray}
with $s,t,u$ the Mandelstam kinematical variables and $\nu=(s-u)/2$. The Regge amplitudes and the relevant phenomenological inputs are given in details in Refs.~\cite{bargerbook,RuizdeElvira:2010cs}.


The general linear relations between the $t$- and $s$-channel isospin amplitudes can be found in many text books in literature, e.g. the one in Ref.~\cite{collinsbook}. Usually the averaging integration region $\nu_2 -\nu_1$ in Eq.~\eqref{eq.defsd} should be taken as $k\cdot{\rm GeV}^2$, with $k=1,2,3,\cdots$. The semilocal duality in Eq.~\eqref{eq.defsd} should in principle work well for the forward scattering amplitude, i.e. by taking $t=0$. It is reasonable to also consider small changes of the $t$, e.g., by taking $t=t_{\rm th}=4m_\pi^2$ for the $\pi\pi$ scattering. Regarding the exponent $n$ in Eq.~\eqref{eq.defsd}, clearly different values of $n$ allow us to probe the interactions in different energy ranges. We will study the duality by setting $n$ at several integers from 0 to 3, which are demonstrated to be proper for the interested energy region below 3~GeV$^2$~\cite{RuizdeElvira:2010cs}. In practice it turns out to be useful to consider the ratios of the average integrations for the $\pi\pi$ scattering, e.g., two different types of ratios are considered in Refs.~\cite{RuizdeElvira:2010cs,Guo:2012ym,Guo:2012yt} 
\begin{eqnarray}\label{eq.defsdr}
 R^I_n &=& \frac{ \int_{\nu_1}^{\nu_2} \nu^{-n}\, {\rm Im}\, T_{\rm t,Hadron}^{(I)}(\nu, t)\, d\nu}
{\int_{\nu_1}^{\nu_3} \nu^{-n}\, {\rm Im}\, T_{\rm t,Hadron}^{(I)}(\nu, t)\, d\nu}\,, \\ 
 F_n^{I I'} &=& \frac{ \int_{\nu_1}^{\nu_{\rm max} } \nu^{-n}\, {\rm Im}\, T_{\rm t,Hadron}^{(I)}(\nu, t)\, d\nu}
{\int_{\nu_1}^{\nu_{\rm max}} \nu^{-n}\, {\rm Im}\, T_{\rm t,Hadron}^{(I')}(\nu, t)\, d\nu}\,,\label{eq.defsdf}
\end{eqnarray}
where $\nu_1, \nu_2$ and $\nu_3$ will be set at the $\pi\pi$ threshold, 1~GeV$^2$ and 2~GeV$^2$, respectively. Two different values for $\nu_{\rm max}=$ 1 or 2 GeV~$^2$ are tested in our study. Nevertheless, since we include the excited scalar resonances around $1.4$~GeV, it is meaningful to fix $\nu_{\rm max}=2$~GeV$^2$, instead of 1~GeV$^2$. Indeed the results with $\nu_{\rm max}=2$~GeV$^2$ turn out to be more reasonable than the case with $\nu_{\rm max}=1$~GeV$^2$. Therefore in the following discussions, we only show the results by taking  $\nu_{\rm max}=2$~GeV$^2$. The integrands in the ratios~\eqref{eq.defsdr} and \eqref{eq.defsdf} can be decomposed into a set of the sum of the partial-wave amplitudes~\cite{RuizdeElvira:2010cs,Guo:2012yt}, which can be calculated with unitarized chiral approach in Eq.~\eqref{eq.ut}. In this way one can discern the roles of the resonances played in the semilocal duality. 

We focus on the situation of the fulfillment of semilocal duality for the $\pi\pi$ scattering here. The values for the ratios with different isospin numbers and $n$ are summarized in Table~\ref{tab:sdratios}. The smaller(larger) values of $n$ enable us to probe the fulfillment of the semilocal duality in the higher(lower) energy region. Apart from the scalar and vector resonances, it is found that the inclusion of the $D$-wave tensor resonances generally improves the fulfillment of the semilocal duality, except the $n=0$ case. To incorporate the $D$-wave contributions to the ratios~\eqref{eq.defsdr} and \eqref{eq.defsdf}, we follow Ref.~\cite{Ecker:2007us} to include the tree-level tensor exchanges. In addition, the other contributions to the $D$-wave amplitudes from the chiral loops, higher order contact terms and the crossed-channel scalar and vector exchanges are also taken into account. The additional parameters related to the tensor resonances are fixed by properly reproducing the $f_2(1270)$ pole~\cite{Guo:2012yt}. The ratios of $F_n^{20}$ and $F_n^{21}$ are particularly interesting to probe the semilocal duality, since the Regge contributions to the $t$-channel amplitudes with $I=2$ are greatly suppressed. As a result, the values of $F_n^{20}$ and $F_n^{21}$ should tend to zero according to the Regge theory. Therefore one would expect the cancellations between the scalar, vector and tensor resonance exchanges for the integrals with $I=2$ in the ratios of $F_n^{20}$ and $F_n^{21}$. By only including the scalar or vector resonance contributions, one obtains that the magnitude of $F_n^{21}$ should approach to 1, which can be considered as a value that signals the complete violation of the semilocal duality. For the $n=1,2,3$ cases, the magnitudes of the $F_n^{21}$ are smaller than 0.3, which indicate that the semilocal duality is well satisfied. Similar conclusion is also obtained for the $F_n^{20}$ case. Regarding the values of the $R_n^{I}$, generally speaking the semilocal duality is better satisfied for higher values of $n$ and we verify that the inclusion of the $D$-wave tensor resonance contributions plays relevant roles in the description of the semilocal duality. It is verified that the results when taking $t=0$ lead to quantitatively similar conclusions as the case with $t=4m_\pi^2$.

\begin{table}[ht]
\renewcommand{\tabcolsep}{0.05cm}
\renewcommand{\arraystretch}{1.2}
\begin{center}
{\small 
\begin{tabular}{l|l|llllll}
\hline
 n &  &&  $R_n^0$    && $R_n^1$   && $F_n^{21}$  
\\
[0.1cm] 
\hline
\hline
Regge & 0 && 0.225  &&  0.325  &&   $\simeq 0$  
\\[0.1cm] 
 &  1     && 0.425  &&  0.578  &&   $\simeq 0$  
\\[0.1cm] 
 & 2     && 0.705 &&  0.839  &&   $\simeq 0$     
\\[0.1cm] 
 &  3    && 0.916  &&  0.966  &&   $\simeq 0$  
 \\[0.1cm]   
\hline 
Hadrons    & 0   && 0.410  &&  0.453   &&  0.531      
\\[0.1cm] 
{\footnotesize $(S+P+D)$} & 1 && 0.653  &&  0.694    &&  0.154   
\\[0.1cm] 
           & 2   && 0.850  &&  0.875   &&  0.027      
\\[0.1cm] 
           & 3   && 0.954  &&  0.965   &&  0.225  
\\[0.1cm]  
\hline 
\hline
\end{tabular}
}
\caption{The results of the ratios defined in Eqs.~\eqref{eq.defsdr} and \eqref{eq.defsdf} to quantify the semilocal duality in the $\pi\pi$ scattering. All the numbers shown in this table are evaluated by taking $t=4m_\pi^2$ and $\nu_{\rm max}=2$~GeV$^2$. The entries in the hadronic parts include the contributions from the $S$, $P$ and $D$ waves. } 
\label{tab:sdratios}
\end{center}
\end{table}

\section{$N_C$ behaviors of the resonances and various physical quantities}

\begin{figure}[htbp]
\begin{center}
\includegraphics[angle=0, width=0.95\textwidth]{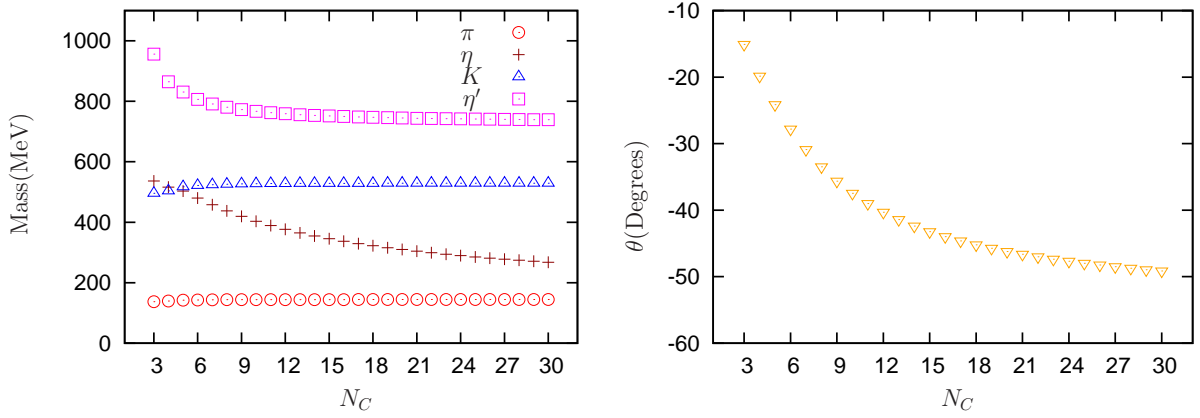}  
\caption{ The $N_C$ evolutions for the masses of the pNGBs (left panel) and the LO $\eta$-$\eta'$ mixing angle (right panel).} \label{fig.ncpsmassandtheta}
\end{center}
\end{figure}

\begin{figure}[htbp]
\begin{center}
\includegraphics[angle=0, width=0.95\textwidth]{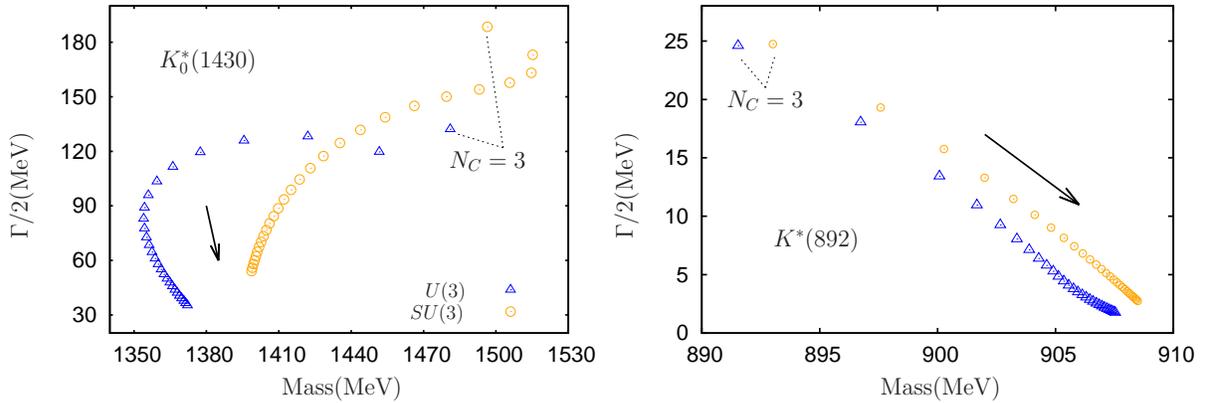}  
\caption{ Comparisons of the $N_C$ trajectories from the $U(3)$ and $SU(3)$ cases. } \label{fig.ncpoleu3su3}
\end{center}
\end{figure}

As mentioned previously, the $U(3)$ chiral theory provides a more appropriate theoretical framework to study the large $N_C$ dynamics of QCD than the $SU(3)$ case, since the singlet $\eta_0$ would become the ninth pNGB in the large $N_C$ and chiral limits. Indeed rather different $N_C$ evolutions for the masses of the $\pi, K$ and $\eta$ in the $U(3)$ chiral theory, which can be seen in Fig.~\ref{fig.ncpsmassandtheta}, have been found, when compared to the more or less flat behaviors in the $SU(3)$ case. The most significant change happens for the $\eta$ meson, which mass greatly decreases to 300~MeV and tends to become degenerate with the pion. For the kaon and $\eta'$, their masses still remain large in the large $N_C$ mainly due to the large physical strange quark mass. The leading order $\eta$-$\eta'$ mixing angle $\theta$ monotonically decreases and approaches to the ideal mixing value in the large $N_C$ limit. The $N_C$ evolutions of the masses for the pNGBs and the $\eta$-$\eta'$ mixing will be taken into account in the following discussions on the $N_C$ behaviors of the resonances and various physical quantities, including the spectral functions integrals and the ratios defined in the previous section to quantify the semilocal duality. It is mentioned that previous works on the study of $N_C$ behaviors of the resonances~\cite{nctrajmeson} have neglected the QCD $\uao$ anomaly effect. In order to explicitly show the influences of the QCD $\uao$ effect in the determinations of the $N_C$ trajectories for the resonances, we propose a way to imitate the $SU(3)$ results from the $U(3)$ case, which include the following operations: fixing the leading order mixing angle $\theta$ at zero throughout, freezing the $\pi, K$ and $\eta$ masses at their physical values and fixing the $\eta'$ mass at its leading order value. The results can be seen in Fig.~\ref{fig.ncpoleu3su3}, where the resonance poles of the $K_0^{*}(1430)$ and $K^*(892)$ are taken as examples to illustrate the differences from the $U(3)$ and $SU(3)$ chiral theories. Generally we can conclude that the QCD $\uao$ effects are not negligible and they have more important influences on the scalar resonances than the vector ones.

\begin{figure}[htbp]
\begin{center}
\includegraphics[angle=0, width=0.95\textwidth]{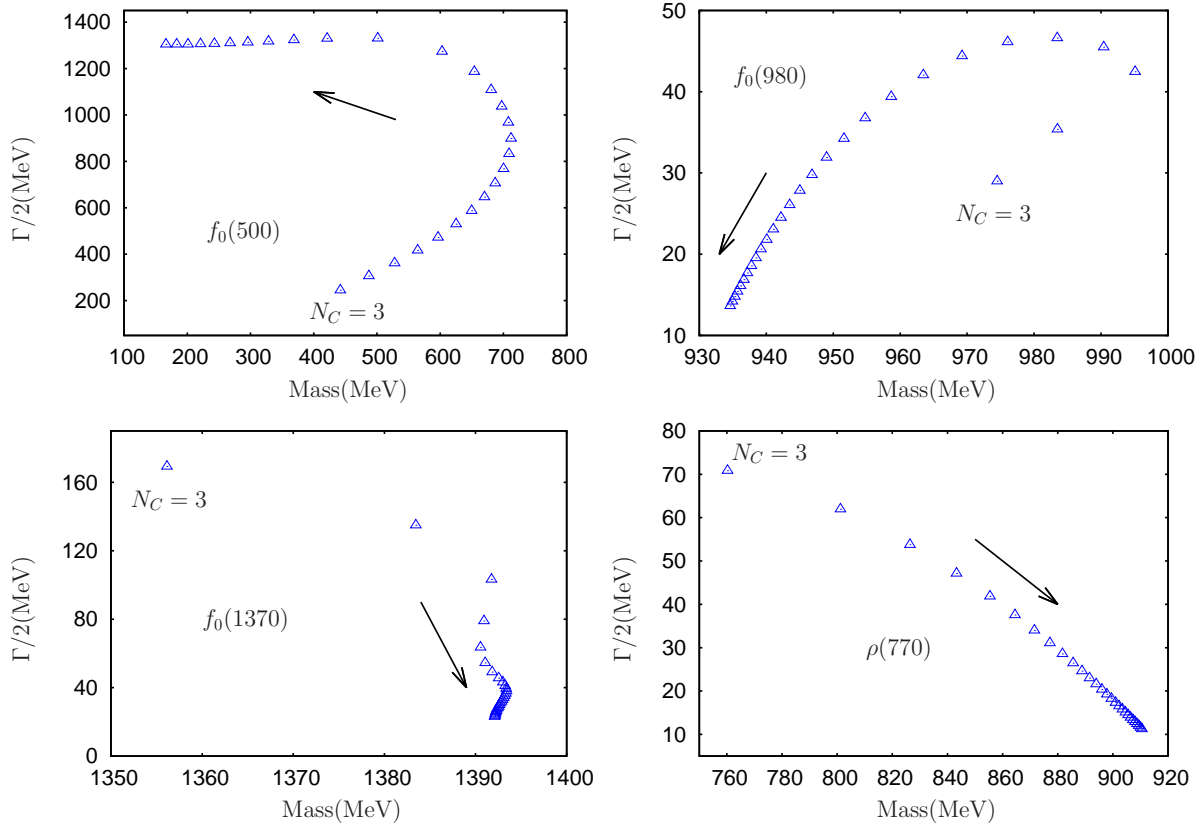}  
\caption{ $N_C$ trajectories of the scalar and vector resonance poles in the $\pi\pi$ scattering.} \label{fig.ncpole}
\end{center}
\end{figure}

According to the large $N_C$ QCD~\cite{largenc}, the mass and width for the conventional $\bar{q}q$ meson when $N_C \to \infty$ scale as $N_C^0$ and $1/N_C$, respectively. This fact provides us a qualitative criteria to study the inner structures of hadrons, once the $N_C$ trajectories of the resonance poles are known. In order to obtain the $N_C$ behaviors of the resonance poles, we need to provide the $N_C$ scaling rules for the various parameters in the unitarized scattering amplitudes. In Refs.~\cite{Guo:2012ym,Guo:2012yt} detailed discussions on the leading and subleading $N_C$ scalings of the relevant parameters are given. By taking into account the $N_C$ evolutions of the pNGBs' masses, the $\eta$-$\eta'$ mixing angle and the various parameters, the corresponding results for the $N_C$ trajectories of the scalar and vector resonances appearing in the $\pi\pi$ scattering are shown in Fig.~\ref{fig.ncpole}. The most important lesson we learn is that the poles of $f_0(980)$, $f_0(1370)$ and $\rho(770)$ fall down to the real axis when $N_C$ becomes large, indicating that at least there are some $\bar{q}q$ seeds inside these resonances in the large $N_C$ limit, which are also supported in our later calculations by using the generalized Weinberg compositeness relations in Refs.~\cite{Guo:2015daa,Gao:2018jhk}. To be more specific, it is verified that the width of the $\rho(770)$ behaves perfectly as $1/N_C$ for $N_C\geq3$ and its mass approaches to a constant for $N_C\geq 15$. Therefore the $N_C$ trajectory of the $\rho(770)$ clearly manifests itself as a standard $\bar{q}q$ resonance. For the $f_0(980)$ and $f_0(1370)$, although their $N_C$ trajectories show some peculiar trends when $N_C\leq 10$, the widths of both resonances approach to zero and their masses tend to constants for large values of $N_C$, implying that important $\bar{q}q$ components start to become dominant when $N_C\to \infty$. While for the $f_0(500)$ pole, its $N_C$ trajectory tends to go deep in the complex plane, instead of falling down to the real axis, i.e. the width of the $f_0(500)$ obviously does not behave as $1/N_C$ even at large $N_C$. This tells us that in our study the $\bar{q}q$ component does not seem playing dominant roles in the $f_0(500)$ when $N_C \to \infty$.

\begin{figure}[htbp]
\begin{center}
\includegraphics[angle=0, width=0.99\textwidth]{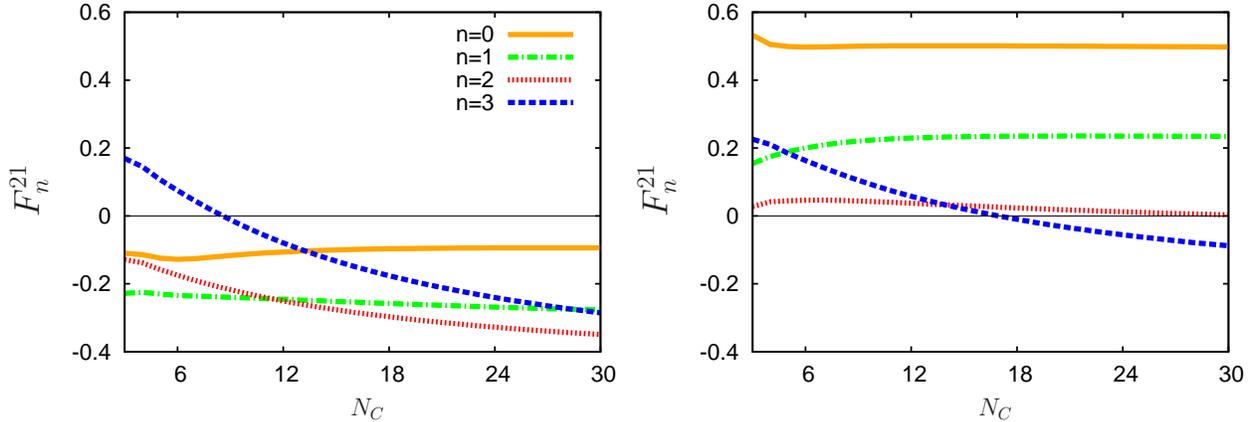}  
\caption{ The $N_C$ evolutions of the ratios $F_n^{21}$ defined in Eq.~\eqref{eq.defsdf}. The curves correspond to the results by including the subleading $N_C$ scalings of the various parameters~\cite{Guo:2012yt}. The left panel is to show the results by only considering the $S$ and $P$ waves. The right panel includes the contributions from the $D$ wave with tensor resonances, as well as the $S$ and $P$ waves.   } \label{fig.ncf21}
\end{center}
\end{figure}

The complex $N_C$ behaviors of the various resonances will be also reflected in the $N_C$ evolutions of the physical quantities, such as the form factors, spectral functions and semilocal dualities, etc. Regarding the $N_C$ evolutions of the spectral function sum rules, it is explicitly verified that the spectral integrals in Eq.~\eqref{eq.defw} obtained in the chiral limit, perfectly scale as $N_C$, as expected from the large $N_C$ QCD~\cite{Guo:2012ym,Guo:2012yt}. It is found that the subtle subleading $N_C$ scalings of the various parameters in the unitarized amplitudes also play relevant roles in the $N_C$ study of the semilocal duality. We show in Fig.~\ref{fig.ncf21} the $N_C$ evolutions of the ratios \eqref{eq.defsdf} by including the subleading $N_C$ scalings of the various parameters~\cite{Guo:2012yt}. The Regge theory predicts the vanishing values of the ratios $F_n^{21}$. Indeed the magnitudes of the $F_n^{21}$ ratios turn to be small for a wide range of $N_C$, indicating that there are cancellations between the contributions from different types of resonances. Nevertheless the cancellation pattern is not universal at different values of $N_C$. We find that rather different cancellation patterns happen for different values of $N_C$. E.g., in the physical case with $N_C=3$ the bump around the $\sigma$ resonance region gives dominant contribution that balances the one from the $\rho(770)$ in the ratio $F_2^{21}$. When $N_C=30$ the bump in the $\sigma$ region barely contributes any more and the contribution from the $\rho(770)$ to the $F_2^{21}$ also get greatly reduced. For the ratio  $F_0^{21}$, the $\rho(770)$ contribution for all the values of $N_C$ is mainly canceled by the one from the $f_0(1370)$, which however barely affects the ratios with $n\ge 1$. Comparing with the left and right panels of Fig.~\ref{fig.ncf21}, we see that the $D$-wave tensor contributions generally improve the fulfillment of the semilocal duality, except the $n=0$ case. This indicates that more types of resonances would be needed to better fulfill the duality for the $n=0$ ratio.


\section{Summary and conclusions}

Recently we have continuously pushed forward the consistent higher-order calculations in the $U(3)$ chiral effective field theory, which include not only the meson-meson scattering and the form factors briefly discussed here, but also the systematical calculations of the light pseudoscalar meson properties, the $\eta$-$\eta'$ mixing and their thermal behaviors. The $U(3)$ chiral effective field theory can be clearly used to investigate more subjects on the hadron phenomenologies than the $SU(3)$ case. The energy regions covered by the $U(3)$ theory in principle are higher than the $SU(3)$ one. It has been also demonstrated to be useful and efficient to analyze the various lattice data~\cite{Guo:2015xva,Guo:2016zep,Gu:2018swy}. 

In this note we mainly focus on the resonance dynamics in various physical quantities calculated in the $U(3)$ resonance chiral theory. Special attention has been paid to the $N_C$ evolutions of the resonance poles and various physical quantities, including the spectral functions and the ratios to quantify the semilocal duality. To properly take into account the nonperturbative strong interactions between the meson pairs, the approximated $N/D$ method is employed to unitarize the perturbative $U(3)$ chiral amplitudes and the scalar form factors. All the relevant scalar resonances below around $1.4$~GeV and almost all the ground vector resonances are obtained in a consistent theoretical framework in our study, including $f_0(500)$, $f_0(980)$,  $f_0(1370)$, $K_0^*(800)$, $K_0^*(1430)$, $a_0(980)$, $a_0(1450)$, $\rho(770)$, $K^*(892)$ and $\phi(1020)$. The scalar spectral functions of the two-point correlators are calculated in terms of the unitarized scalar form factors and the pseudoscalar spectral functions are approximated by the single-meson contributions. Two kinds of ratios have been defined to quantify the semilocal duality, which gives us further insight into the underlying relations between different types of resonances. 
The $N_C$ trajectory curves of the resonance poles reveal that in the large $N_C$ the $\bar{q}q$ components seem playing marginal roles for the $f_0(500)$, $a_0(980)$ and $K_0^*(800)$, while there are strong evidences in our study that the  $\bar{q}q$ seeds are important for the other resonances mentioned above. The scalar and pseudoscalar spectral sum rules are found to be rather well satisfied and perfectly scale as $N_C$ in the chiral limit. Although to a good extent the semilocal duality is generally fulfilled, the cancellation patterns among the contributions from the different resonances are found to be subtle at different values of $N_C$. We foresee that the $U(3)$ chiral theory can be further investigated to address many interesting phenomenological and lattice problems.

\section*{Acknowledgement}
I thank J.~A.~Oller and J.~Ruiz de Elvira for the pleasant collaborations and useful discussions on the topics covered here. 
This work is supported in part by the Natural Science Foundation of China(NSFC) under Grants Nos.~11975090 and 11575052, the Natural Science Foundation of Hebei Province under Contract No.~A2015205205.


\begin{thebibliography}{90}

\bibitem{bibchptew}
A.~Dobado and J.~Pelaez,
Nucl. Phys. B \textbf{425}, 110-136 (1994)
doi:10.1016/0550-3213(94)90174-0
[arXiv:hep-ph/9401202 [hep-ph]].

J.~Oller,
Phys. Lett. B \textbf{477}, 187-194 (2000)
doi:10.1016/S0370-2693(00)00185-4
[arXiv:hep-ph/9908493 [hep-ph]].

F.~K.~Guo, P.~Ruiz-Femenia and J.~J.~Sanz-Cillero,
Phys. Rev. D \textbf{92}, 074005 (2015)
doi:10.1103/PhysRevD.92.074005
[arXiv:1506.04204 [hep-ph]].

A.~Dobado, F.~J.~Llanes-Estrada and J.~J.~Sanz-Cillero,
JHEP \textbf{03}, 159 (2018)
doi:10.1007/JHEP03(2018)159
[arXiv:1711.10310 [hep-ph]].

O.~Cata and C.~Muller,
Nucl. Phys. B \textbf{952}, 114938 (2020)
doi:10.1016/j.nuclphysb.2020.114938
[arXiv:1906.01879 [hep-ph]].

C.~Krause, A.~Pich, I.~Rosell, J.~Santos and J.~J.~Sanz-Cillero,
JHEP \textbf{05}, 092 (2019)
doi:10.1007/JHEP05(2019)092
[arXiv:1810.10544 [hep-ph]].

\bibitem{Weinberg:1978kz}
S.~Weinberg,
Physica A \textbf{96}, no.1-2, 327-340 (1979)
doi:10.1016/0378-4371(79)90223-1

  
\bibitem{Gasser:1983yg} 
  J.~Gasser and H.~Leutwyler,
  Annals Phys.\  {\bf 158}, 142 (1984).
  doi:10.1016/0003-4916(84)90242-2
  
  
\bibitem{Gasser:1984gg} 
  J.~Gasser and H.~Leutwyler,
  Nucl.\ Phys.\ B {\bf 250}, 465 (1985).
  doi:10.1016/0550-3213(85)90492-4

\bibitem{largenc}
G.~'t Hooft,
Nucl. Phys. B \textbf{72}, 461 (1974)
doi:10.1016/0550-3213(74)90154-0; 

G.~'t Hooft,
Nucl. Phys. B \textbf{75}, 461-470 (1974)
doi:10.1016/0550-3213(74)90088-1; 


E.~Witten,
Nucl. Phys. B \textbf{160}, 57-115 (1979)
doi:10.1016/0550-3213(79)90232-3

\bibitem{ua1anomaly}
S.~L.~Adler and W.~A.~Bardeen,
Phys. Rev. \textbf{182}, 1517-1536 (1969)
doi:10.1103/PhysRev.182.1517

W.~A.~Bardeen,
Phys. Rev. \textbf{184}, 1848-1857 (1969)
doi:10.1103/PhysRev.184.1848


K.~Fujikawa,
Phys. Rev. D \textbf{21}, 2848 (1980)
doi:10.1103/PhysRevD.21.2848

\bibitem{ua1nc}

E.~Witten,
Nucl. Phys. B \textbf{156}, 269-283 (1979)
doi:10.1016/0550-3213(79)90031-2;

S.~R.~Coleman and E.~Witten,
Phys. Rev. Lett. \textbf{45}, 100 (1980)
doi:10.1103/PhysRevLett.45.100;

G.~Veneziano,
Nucl. Phys. B \textbf{159}, 213-224 (1979)
doi:10.1016/0550-3213(79)90332-8

\bibitem{HerreraSiklody:1996pm}
  P.~Herrera-Siklody, J.~I.~Latorre, P.~Pascual and J.~Taron,
  Nucl.\ Phys.\ B {\bf 497}, 345 (1997)
  doi:10.1016/S0550-3213(97)00260-5
  [hep-ph/9610549].


\bibitem{Kaiser:2000gs}
  R.~Kaiser and H.~Leutwyler,
  Eur.\ Phys.\ J.\ C {\bf 17}, 623 (2000)
  doi:10.1007/s100520000499
  [hep-ph/0007101].


\bibitem{Guo:2011pa}
Z.~Guo and J.~Oller,
Phys. Rev. D \textbf{84}, 034005 (2011)
doi:10.1103/PhysRevD.84.034005
[arXiv:1104.2849 [hep-ph]].

\bibitem{Guo:2012ym}
Z.~Guo, J.~Oller and J.~Ruiz de Elvira,
Phys. Lett. B \textbf{712}, 407-412 (2012)
doi:10.1016/j.physletb.2012.05.021
[arXiv:1203.4381 [hep-ph]].

\bibitem{Guo:2012yt}
Z.~Guo, J.~Oller and J.~Ruiz de Elvira,
Phys. Rev. D \textbf{86}, 054006 (2012)
doi:10.1103/PhysRevD.86.054006
[arXiv:1206.4163 [hep-ph]].




\bibitem{Guo:2016zep}
Z.~Guo, L.~Liu, U.~Mei\ss ner, J.~Oller and A.~Rusetsky,
Phys. Rev. D \textbf{95}, no.5, 054004 (2017)
doi:10.1103/PhysRevD.95.054004
[arXiv:1609.08096 [hep-ph]].
  
\bibitem{Ecker:1988te}
G.~Ecker, J.~Gasser, A.~Pich and E.~de Rafael,
Nucl. Phys. B \textbf{321}, 311-342 (1989)
doi:10.1016/0550-3213(89)90346-5


\bibitem{rxtpheno}
D.~Dumm, P.~Roig, A.~Pich and J.~Portoles,
Phys. Lett. B \textbf{685}, 158-164 (2010)
doi:10.1016/j.physletb.2010.01.059
[arXiv:0911.4436 [hep-ph]].

P.~Roig, A.~Guevara and G.~Lopez Castro,
Phys. Rev. D \textbf{89}, no.7, 073016 (2014)
doi:10.1103/PhysRevD.89.073016
[arXiv:1401.4099 [hep-ph]].

Z.~H.~Guo and P.~Roig,
Phys. Rev. D \textbf{82}, 113016 (2010)
doi:10.1103/PhysRevD.82.113016
[arXiv:1009.2542 [hep-ph]].

Z.~H.~Guo,
Phys. Rev. D \textbf{78}, 033004 (2008)
doi:10.1103/PhysRevD.78.033004
[arXiv:0806.4322 [hep-ph]].

R.~Escribano, P.~Masjuan and J.~J.~Sanz-Cillero,
JHEP \textbf{05}, 094 (2011)
doi:10.1007/JHEP05(2011)094
[arXiv:1011.5884 [hep-ph]].

M.~Albaladejo and J.~Oller,
Phys. Rev. Lett. \textbf{101}, 252002 (2008)
doi:10.1103/PhysRevLett.101.252002
[arXiv:0801.4929 [hep-ph]].

R.~Escribano, S.~Gonzalez-Solis, M.~Jamin and P.~Roig,
JHEP \textbf{09}, 042 (2014)
doi:10.1007/JHEP09(2014)042
[arXiv:1407.6590 [hep-ph]].

P.~Roig and P.~Sanchez-Puertas,
Phys. Rev. D \textbf{101}, no.7, 074019 (2020)
doi:10.1103/PhysRevD.101.074019
[arXiv:1910.02881 [hep-ph]].


\bibitem{Cirigliano:2006hb}
V.~Cirigliano, G.~Ecker, M.~Eidemuller, R.~Kaiser, A.~Pich and J.~Portoles,
Nucl. Phys. B \textbf{753}, 139-177 (2006)
doi:10.1016/j.nuclphysb.2006.07.010
[arXiv:hep-ph/0603205 [hep-ph]].


\bibitem{Beisert:2001qb}
N.~Beisert and B.~Borasoy,
Eur. Phys. J. A \textbf{11}, 329-339 (2001)
doi:10.1007/s100500170072
[arXiv:hep-ph/0107175 [hep-ph]].


\bibitem{Beisert:2002ad}
N.~Beisert and B.~Borasoy,
Nucl. Phys. A \textbf{705}, 433-454 (2002)
doi:10.1016/S0375-9474(02)00652-8
[arXiv:hep-ph/0201289 [hep-ph]].

\bibitem{Oller:2019opk}
J.~Oller,
Prog. Part. Nucl. Phys.110, 103728(2020) 
doi:10.1016/j.ppnp.2019.103728
[arXiv:1909.00370 [hep-ph]].

\bibitem{Oller:2020guq}
J.~Oller,
[arXiv:2005.14417 [hep-ph]].

\bibitem{Oller:1998zr}
J.~Oller and E.~Oset,
Phys. Rev. D \textbf{60}, 074023 (1999)
doi:10.1103/PhysRevD.60.074023
[arXiv:hep-ph/9809337 [hep-ph]].


\bibitem{Oller:2000fj}
J.~Oller and U.~G.~Meissner,
Phys. Lett. B \textbf{500}, 263-272 (2001)
doi:10.1016/S0370-2693(01)00078-8
[arXiv:hep-ph/0011146 [hep-ph]].


\bibitem{Dudek:2016cru}
J.~J.~Dudek \textit{et al.} [Hadron Spectrum],
Phys. Rev. D \textbf{93}, no.9, 094506 (2016)
doi:10.1103/PhysRevD.93.094506
[arXiv:1602.05122 [hep-ph]].

\bibitem{speclsumrule}

C.~W.~Bernard, A.~Duncan, J.~LoSecco and S.~Weinberg,
Phys. Rev. D \textbf{12}, 792 (1975)
doi:10.1103/PhysRevD.12.792;

H.~Leutwyler,
Nucl. Phys. B \textbf{337}, 108-118 (1990)
doi:10.1016/0550-3213(90)90253-A; 

M.~F.~Golterman and S.~Peris,
Phys. Rev. D \textbf{61}, 034018 (2000)
doi:10.1103/PhysRevD.61.034018
[arXiv:hep-ph/9908252 [hep-ph]];

B.~Moussallam,
Eur. Phys. J. C \textbf{14}, 111-122 (2000)
doi:10.1007/s100520050738
[arXiv:hep-ph/9909292 [hep-ph]]; 

B.~Moussallam,
JHEP \textbf{08}, 005 (2000)
doi:10.1088/1126-6708/2000/08/005
[arXiv:hep-ph/0005245 [hep-ph]];

M.~Jamin and M.~Munz,
Z. Phys. C \textbf{60}, 569-578 (1993)
doi:10.1007/BF01560056
[arXiv:hep-ph/9208201 [hep-ph]].

\bibitem{bargerbook}
V.~D.~Barger and D.~B.~Cline, {\it Phenomenological Theoriesof High Energy Scattering} (W.~A.~Benjamin, New York, 1969). 

\bibitem{RuizdeElvira:2010cs}
J.~Ruiz de Elvira, J.~Pelaez, M.~Pennington and D.~Wilson,
Phys. Rev. D \textbf{84}, 096006 (2011)
doi:10.1103/PhysRevD.84.096006
[arXiv:1009.6204 [hep-ph]].

\bibitem{collinsbook} P.D.B. Collins, {\it An introduction to Regge theory and high energy physics}.

\bibitem{Ecker:2007us}
G.~Ecker and C.~Zauner,
Eur. Phys. J. C \textbf{52}, 315-323 (2007)
doi:10.1140/epjc/s10052-007-0372-x
[arXiv:0705.0624 [hep-ph]].

\bibitem{nctrajmeson}
J.~Pelaez,
Phys. Rev. Lett. \textbf{92}, 102001 (2004)
doi:10.1103/PhysRevLett.92.102001
[arXiv:hep-ph/0309292 [hep-ph]].

J.~Pelaez,
Mod. Phys. Lett. A \textbf{19}, 2879-2894 (2004)
doi:10.1142/S0217732304016160
[arXiv:hep-ph/0411107 [hep-ph]].
 
J.~Pelaez and G.~Rios,
Phys. Rev. Lett. \textbf{97}, 242002 (2006)
doi:10.1103/PhysRevLett.97.242002
[arXiv:hep-ph/0610397 [hep-ph]].

Z.~Sun, L.~Xiao, Z.~Xiao and H.~Zheng,
Mod. Phys. Lett. A \textbf{22}, 711-718 (2007)
doi:10.1142/S0217732307023304
[arXiv:hep-ph/0503195 [hep-ph]].

Z.~H.~Guo, L.~Xiao and H.~Zheng,
Int. J. Mod. Phys. A \textbf{22}, 4603-4616 (2007)
doi:10.1142/S0217751X0703710X
[arXiv:hep-ph/0610434 [hep-ph]].

Z.~Guo, J.~Sanz Cillero and H.~Zheng,
JHEP \textbf{06}, 030 (2007)
doi:10.1088/1126-6708/2007/06/030
[arXiv:hep-ph/0701232 [hep-ph]].


J.~Nieves and E.~Ruiz Arriola,
Phys. Rev. D \textbf{80}, 045023 (2009)
doi:10.1103/PhysRevD.80.045023
[arXiv:0904.4344 [hep-ph]].

J.~Nieves, A.~Pich and E.~Ruiz Arriola,
Phys. Rev. D \textbf{84}, 096002 (2011)
doi:10.1103/PhysRevD.84.096002
[arXiv:1107.3247 [hep-ph]].


T.~Ledwig, J.~Nieves, A.~Pich, E.~Ruiz Arriola and J.~Ruiz de Elvira,
Phys. Rev. D \textbf{90}, no.11, 114020 (2014)
doi:10.1103/PhysRevD.90.114020
[arXiv:1407.3750 [hep-ph]].

Z.~Y.~Zhou and Z.~Xiao,
Phys. Rev. D \textbf{83}, 014010 (2011)
doi:10.1103/PhysRevD.83.014010
[arXiv:1007.2072 [hep-ph]].

L.~Dai, X.~Wang and H.~Zheng,
Commun. Theor. Phys. \textbf{57}, 841-848 (2012)
doi:10.1088/0253-6102/57/5/15
[arXiv:1108.1451 [hep-ph]].

L.~Y.~Dai, X.~G.~Wang and H.~Q.~Zheng,
Commun. Theor. Phys. \textbf{58}, 410-414 (2012)
doi:10.1088/0253-6102/58/3/15
[arXiv:1206.5481 [hep-ph]].

L.~Y.~Dai and U.~G.~Mei\ss ner,
Phys. Lett. B \textbf{783}, 294-300 (2018)
doi:10.1016/j.physletb.2018.06.071
[arXiv:1706.10123 [hep-ph]].

T.~Wolkanowski, M.~Soltysiak and F.~Giacosa,
Nucl. Phys. B \textbf{909}, 418-428 (2016)
doi:10.1016/j.nuclphysb.2016.05.025
[arXiv:1512.01071 [hep-ph]].

\bibitem{Guo:2015daa}
Z.~H.~Guo and J.~Oller,
Phys. Rev. D \textbf{93}, no.9, 096001 (2016)
doi:10.1103/PhysRevD.93.096001
[arXiv:1508.06400 [hep-ph]].


\bibitem{Gao:2018jhk}
R.~Gao, Z.~H.~Guo, X.~W.~Kang and J.~Oller,
Adv. High Energy Phys. \textbf{2019}, 4651908 (2019)
doi:10.1155/2019/4651908
[arXiv:1812.07323 [hep-ph]].


\bibitem{Guo:2015xva}
X.~Guo, Z.~Guo, J.~A.~Oller and J.~J.~Sanz-Cillero,
JHEP \textbf{06}, 175 (2015)
doi:10.1007/JHEP06(2015)175
[arXiv:1503.02248 [hep-ph]].

\bibitem{Gu:2018swy}
X.~Gu, C.~Duan and Z.~Guo,
Phys. Rev. D \textbf{98}, no.3, 034007 (2018)
doi:10.1103/PhysRevD.98.034007
[arXiv:1803.07284 [hep-ph]].


\end{thebibliography}
\end{document}